\documentclass{article}
\usepackage{amssymb}
\usepackage{amsmath,bm}
\usepackage{array}
\usepackage{caption,graphicx}

\newcommand{\be}{\begin{equation}}
\newcommand{\ee}{\end{equation}}

\def\un{{\rm 1\mkern-4mu I}}

\title{The lure of conformal symmetry\footnote{Invited talk at the conference ``Conformal Invariance and Harmonic Analysis'', Institut de Physique Th\'eorique de Saclay, Gif-sur-Yvette, 5 December 2018. Bures-sur-Yvette preprint, IHES/P/19/01.}}
\author{Ivan Todorov}
\date{
\small Institut des Hautes \'Etudes Scientifiques, 35 route de Chartres, 
\\ F-91440 Bures-sur-Yvette, France;
\\ and
\\ Institute for Nuclear Research and Nuclear Energy, Bulgarian Academy of Sciences, 
\\ Tsarigradsko Chaussee 72, BG-1784 Sofia, Bulgaria
\\ (permanent address)
\\e-mail address: ivbortodorov@gmail.com
}

\begin{document}

\maketitle

\vglue 2cm

\begin{abstract}
The Clifford algebra ${\rm Cl} (4,1) \simeq {\mathbb C} [4]$, generated by the real (Majorana) $\gamma$-matrices and by a hermitian $\gamma_5$, gives room to the reductive Lie algebra $u(2,2)$ of the conformal group extended by the $u(1)$ helicity operator. Its unitary positive energy ladder representations, constructed by Gerhard Mack and the author 50 years ago, opened the way to a better understanding of zero-mass particles and fields and their relation to the space of bound states of the hydrogen atom. They became a prototypical example of a minimal representation of a non-compact reductive group introduced during the subsequent decade by Joseph.

By the mid 1980's I have developed an analytic approach to compactified Minkowski space $\bar{M}$, suited for extending the notion of vertex operator algebras to higher dimensions. Another 20 years later Nikolay Nikolov and I realized that thermal correlation functions of massless fields on $\bar{M}$ are doubly periodic elliptic functions while the logarithmic derivative of the corresponding partition function is a modular form reproducing Planck's black body radiation law.
\end{abstract}

\newpage

\tableofcontents

%\bigskip

%\noindent {\bf References}

\newpage

Should one aim to emphasize the role of the conformal group in physics, one may start with the early realization -- by Bateman and Cuningham in 1909 -- of the conformal invariance of the classical Maxwell equations and continue with the 1918 Weyl attempt (praised decades later by Dirac) to construct an affine (Weyl) conformal invariant unified field theory$\ldots$ We are choosing a less glorious, rather personal path: from playing with the ${\rm Cl} (3,1)$ $\gamma$-matrices, through the ladder representations of $u(2,2)$ to the modular form $G_4 (\tau)$ and its relation to Planck's black body radiation law. We give in Proposition 3.2 an exponential realization of a quantum  massless field in terms of ladder (creation and annihilation) operators.

\section{The Lie subalgebra $u(2,2)$ of ${\rm Cl} (4,1)$; ladder representation}\label{sec1}

Dirac was recalling in his later years that, on the way of discovering his beautiful equation he was playing with matrices. It is an entertaining game, it can be recommended to curious highschool students. If we endow the Clifford algebra ${\rm Cl} (3,1)$ with a hermitian conjugation such that for
$$
[\gamma_{\mu} , \gamma_{\nu}]_+ = 2 \, \eta_{\mu\nu} \, , \ (\eta_{\mu\nu}) = {\rm diag} (-1,1,1,1) 
$$
we set
\be
\label{eq11}
\gamma^*_{\mu} = \eta_{\mu\mu} \gamma_{\mu} \, , \ (\gamma_{\mu} \gamma_{\nu})^* = \gamma_{\nu}^* \gamma_{\mu}^* \, ,
\ee
then $\gamma_{\mu}$ and their commutators preserve an indefinite hermitian form
\be
\label{eq12}
\widetilde\varphi \, \varphi := \varphi^* \beta \, \varphi \quad \mbox{where} \quad \beta^* = \beta \, , \ \beta^2 = \un \, , \ {\rm tr} \, \beta = 0 \, ,
\ee
so that
\be
\label{eq13}
\gamma_{\mu}^* \, \beta + \beta \, \gamma_{\mu} = 0 \Rightarrow \gamma_{\mu\nu}^* \, \beta + \beta \, \gamma_{\mu\nu} = 0 \quad \mbox{for} \quad \gamma_{\mu\nu} = {\textstyle \frac12} [\gamma_{\mu} , \gamma_{\nu}] \, .
\ee
In fact, ${\rm Cl} (3,1) \simeq {\mathbb R} [4]$ ($\ne {\rm Cl} (1,3)$!) -- that's why we have Majorana spinors -- and the maximal Lie subalgebra of ${\rm Cl} (3,1)$ is $so (3,2) \simeq sp (4,{\mathbb R})$ (spanned by $\gamma_{\lambda}$ and $\gamma_{\mu\nu}$). We need a hermitian chirality matrix $\gamma_5$, anticommuting with $\gamma_{\mu}$, which together with $\gamma_{\mu}$ generates the Clifford algebra ${\rm Cl} (4,1) \simeq {\mathbb C} [4]$, in order to obtain a maximal (16 dimensional) Lie subalgebra $u(2,2)$ of complex $4 \times4$ matrices $X$ satisfying $X^* \beta + \beta X = 0$, thus extending (\ref{eq13}.

\bigskip

\noindent {\bf Remark 1.1.} If we define the {\it Clifford conjugation} $X \to X^+$ as an algebra anti-homomorphism (i.e. such that $(XY)^+ = Y^+ X^+$) with the property $\gamma_a^+ = - \gamma_a$ for all five generators of ${\rm Cl} (4,1)$ then the set of all $X \in {\rm Cl} (4,1)$ for which $X^+ = -X$ spans the Lie algebra $u(2,2)$ -- see \cite{T11}, Proposition 2.2.

\bigskip

We arrive in this way at the conformal Lie algebra $su(2,2) \simeq so (4,2)$ extended by the $u(1)$ helicity operator.

\smallskip

As the conformal group includes dilations it would be broken by dimensional parameters like masses; we can only expect unbroken conformal symmetry for massless particles. But Wigner taught us \cite{W39} that all particles -- massive and massless -- are described by (projective) irreducible representations (IRs) of the (10-dimensional) Poincar\'e subgroup ${\mathcal P}$ of the conformal group. An IR of $U(2,2)$ is expected to split, in general, into a continuum of IRs of ${\mathcal P}$. It turned out that the massless IRs of $U(2,2)$ are exceptional \cite{MT}: they remain irreducible when restricted to ${\mathcal P}$. Such representations were later called {\it minimal} (see \cite{T10} for a review and the {\it Note} after Eq.~(\ref{eq116}) below for more references).

\smallskip

The passage from the 4-dimensional non-unitary representation of $u(2,2)$ to infinite dimensional {\it unitary positive energy irreducible representations} (${\rm UPEIRs}$) uses the creation and annihilation (or emission and absorption) operators of Dirac's oscillators \cite{D27}.

\smallskip

Assume that $\varphi$ and $\widetilde\varphi$ ($= \varphi^* \beta$) obey the bosonic {\it canonical commutation relations:}
\be
\label{eq14}
[\varphi^{\alpha} , \varphi^{\beta}] = 0 = [\widetilde\varphi_{\alpha} , \widetilde\varphi_{\beta}] \, , \ [\varphi^{\alpha} , \widetilde\varphi_{\beta}] = \delta_{\beta}^{\alpha} \quad (\alpha , \beta = 1,2,3,4) \, .
\ee
Then the ``second quantized'' {\it ladder operators} $\widehat X = \widetilde\varphi X \varphi$ satisfy
\be
\label{eq15}
[\widehat X , \widehat Y] = [\widehat{X,Y}] := \widetilde\varphi [X,Y] \varphi \, .
\ee
There are two types of unitary {\it ladder representations} of $u(2,2)$: {\it lowest} and {\it highest weight} representations. To define them we introduce the standard Chevalley-Cartan basis of $su(2,2)$:
\be
\label{eq16}
E_i = \widetilde\varphi_i \, \varphi^{i+1} \, , \ F_i = \widetilde\varphi_{i+1} \, \varphi^i \, , \ H_i = [E_i , F_i] = \widetilde\varphi_i \, \varphi^i - \widetilde\varphi_{i+1} \varphi^{i+1}
\ee
$i = 1,2,3$.

\smallskip

A lowest weight (LW) and a {\it highest weight} (HW) vector are defined as eigenvectors of the Cartan elements $H_i$ annihilated by the lowering and by the raising operators $F_i$ and $E_i$, respectively:
\be
\label{eq17}
F_i \, \vert {\rm LW} \rangle = 0 = (H_{\theta} - 1) \, \vert {\rm LW} \rangle \, , \ E_i \, \vert {\rm HW} \rangle = 0 = (H_{\theta} + 1) \, \vert {\rm HW} \rangle,
\ee
$i=1,2,3$; $\theta$ is the highest root: $H_{\theta} = H_1 + H_2 + H_3$. We shall be interested in LW representations since they are also the positive energy ones. The {\it Fock space vacuum} $\vert 0 \rangle$ ($= \vert 0 \rangle_{\rm LW}$) is in this case the unique LW vector that transforms under an one-dimensional representation of the maximal compact subgroup $S(U(2) \times U(2))$ of $SU(2,2)$:
\be
\label{eq18}
E_1 \vert 0 \rangle = 0 = E_3 \vert 0 \rangle \, , \ (H_c - 2) \, \vert 0 \rangle = 0 \, , \ H_c = H_1 + 2H_2 + H_3
\ee
($H_c$ being the generator of the centre $U(1)$ of $S(U(2) \times U(2))$. We shall demonstrate in Appendix A that the conditions (\ref{eq17}) (i.e. $F_i \vert 0 \rangle = 0$, $i=1,2,3$) and (\ref{eq18}) are equivalent to
\be
\label{eq19}
a \, \vert 0 \rangle = 0 = b \, \vert 0 \rangle \quad (a = (a_1 , a_2) \, , \ b = (b_1 , b_2))
\ee
for
\be
\label{eq110}
a = \frac{1+\beta}2 \, \varphi \, , \ b = \frac{\beta - 1}2 \, \widetilde\varphi \quad \mbox{or} \quad \varphi = \begin{pmatrix} a \\ b^* \end{pmatrix} \, , \ \widetilde\varphi = (a^*,-b)
\ee
(the last two formulas are valid in a $\beta$-diagonal realization of ${\rm Cl} (4,1)$). In order to define the energy-momentum 4-vector one needs the projection on positive eigenvalues of the chirality matrix $\gamma_5$.

\smallskip

The eleven matrices
$$
\gamma_{\mu\nu} = {\textstyle \frac12} \, [\gamma_{\mu} , \gamma_{\nu}] \, , \ \mu , \nu = 0,1,2,3 \, , \ \mu < \nu \, ,
$$
\be
\label{eq111}
\gamma_5 \ {\rm and} \ \gamma_{\mu} \, \Pi_+ = \Pi_- \, \gamma_{\mu} \, , \quad \Pi_{\pm} = \frac{1 \pm \gamma_5}2
\ee
span the Poincar\'e subalgebra of $u(2,2)$ extended by dilations (the automorphism algebra of the Poincar\'e group). We have, $\Pi_+ \, \Pi_- = 0$ so that, in particular,
\be
\label{eq112}
\gamma_{\mu} \, \Pi_+ \, \gamma_{\nu} \, \Pi_+ = 0 \Rightarrow [\gamma_{\mu} \, \Pi_+ , \gamma_{\nu} \, \Pi_+] = 0 \, ; \ \gamma_5 \, \gamma_{\mu} \, \Pi_+ = -\gamma_{\mu} \, \Pi_+ \, ;
\ee
thus $\gamma_{\mu} \, \Pi_+$ satisfy the commutation relations of the generators of translation. As we shall see in Appendix A the 4-momentum $p_{\mu}$ (the hermitian generators of translation) defined by
\be
\label{eq113}
\widetilde\varphi \, \gamma_{\mu} \, \Pi_+ \, \varphi = i \, p_{\mu}
\ee
is massless, i.e. $p^2 = 0$, and has positive energy:
\be
\label{eq114}
p_0 = \sqrt{{\bm p}^2} \geq 0 \quad ({\bm p}^2 = p_1^2 + p_2^2 + p_3^2) \, .
\ee
For fixed (twice) {\it helicity}
\be
\label{eq115}
h =: \widetilde\varphi \, \varphi := \widetilde\varphi \, \varphi - \langle 0 \vert \, \widetilde\varphi \, \varphi \, \vert 0 \rangle = a^* a - b^* b
\ee
the ladder representation of $U(2,2)$ is irreducible and remains irreducible when restricted to the Poincar\'e subgroup (see \cite{MT} and Appendix A below). It is a prominent example of what was later called a {\it minimal representation:} it corresponds to the unique {\it minimal} (6-dimensional) {\it nilpotent orbit} of $su(2,2)$ -- the orbit of the highest root $E_{\theta}$:
$$
\bigl(E_{\theta} = \left[ [ E_1 , E_2 ] , E_3 \right] = a_1^* \, b^*_2 \, , \ [E_1 , E_2] = a_1^* \, b_1^* \, ,
$$
\be
\label{eq116}
[E_2 , E_3] = a_2^* \, b_2^* \, , \ E_1 = a_1^* \, a_2 \, , \ E_3 = -b_1 \, b^*_2 \, , H_{\theta} \bigl) \, .
\ee
The Gelfand-Kirillov dimension of these representations is three -- coinciding with the number of arguments ${\bm p} = (p_1 , p_2 , p_3)$ of the momentum space wave function of a massless particle.

\bigskip

\noindent {\bf Note.} In the mathematical literature the oscillator representation is often referred to Segal-Shale-Weil. If the first papers \cite{S,Sh} were physics' inspired, Weil's work \cite{W} related the subject to number theory and automorphic forms that increased markedly its attraction to mathematicians. This trend was reinforced by another influential paper (of 14 years later) \cite{KV}. (A four-years-old expository paper \cite{AP} emphasizes number theoretic applications and Howe duality; other aspects are covered in \cite{K}.) Minimal representations were introduced by Joseph \cite{J} in mid 1970s. The notion of reductive dual pair (like $U(2,2)$ and its $U(1)$ centre) was introduced by Howe around the same time and eventually published in \cite{H85,H89}. Surveys written by -- and addressed to -- physicists include \cite{GP,T10}.

\bigskip

Coming back to massless particles, the question arises what do we gain by considering a larger (conformal) symmetry group if we end up with ``the same'' Poincar\'e group representation? We shall exploit next (in Sections 2 and 3) two such interrelated features: (i) the existence of a natural conformal compactification $\overline M$ of Minkowski space; (ii) the presence of a complete set of commuting observables, with a discrete spectrum. In particular, the conformal Hamiltonian, whose role was pointed out by Irving Segal \cite{S82},  
\be
\label{eq117}
H = {\textstyle \frac12} \, H_c = {\textstyle \frac12} \, (a^* a + b \, b^*)
\ee
(where $H_c = H_1 + 2H_2 + H_3$, (\ref{eq18})) has a positive integer spectrum in the Fock space ${\mathcal F}_0$ of the zero helicity massless field $\phi$.

%%%
\section{Conformal compactification of space-time and the tube domain}\label{sec2}
\setcounter{equation}{0}

Dirac \cite{D36} has defined conformally compactified Minkowski space as a projective quadric in 6-space:
\be
\label{eq21}
\overline M = Q / {\mathbb R}^* \, , \ Q = \left\{ \xi \in {\mathbb R}^6 \backslash \{0\} : \xi_{\mu} \, \xi^{\mu} := {\bm\xi}^2 - \xi_0^2 = \xi_+ \xi_- \right\} .
\ee
Identifying the cone at infinity $C_{\infty}$ with the intersection $\overline M$ with the hyperplane $\xi_+ = 0$ we can express the rays in $Q \backslash C_{\infty}$ in terms of the coordinates $x^{\mu}$ of a point in Minkowski space $M$ as:
\be
\label{eq22}
Q \backslash C_{\infty} = \{ \xi_{\mu} = \xi_+ x_{\mu} \, , \ \xi_- = \xi_+ x^2 \, ; \ \xi_+ \ne 0 \} \simeq M \, .
\ee
If we denote such an embedding as $x \to \xi_x$ ($y \to \eta_y$) we would have $(\xi_x - \eta_y)^2 = -2 \xi_x \eta_y = \xi_+ \eta_+ (x-y)^2$. The compactified Minkowski space $\overline M$ (\ref{eq21}) is isomorphic to the product of a circle and a 3-sphere with identified opposite points. To see this, it is convenient to use pseudo-orthogonal coordinates $\xi_{-1}$ ($= -\xi^{-1}$) and $\xi_4$ ($= \xi^4$) such that $\xi_{-1}^2 - \xi_4^2 = \xi_+ \xi_-$; then
\be
\label{eq23}
\overline M = Q / {\mathbb R}^* \simeq \{ \xi : \xi_{-1}^2 + \xi_0^2 = 1 = {\bm \xi}^2 + \xi_4^2 \} / \{ \xi \leftrightarrow -\xi \} \, .
\ee
There exists a remarkable complex variable parametrization of $\overline M$ \cite{T86} related to the real coordinates $x \in M$ by a complex conformal transformation $g_c : M \hookrightarrow E_{\mathbb C}$ with no real singularities

$$
\overline M = \left\{ z_{\alpha} = \frac{\xi_{\alpha}}{i \xi_0 - \xi_{-1}} \left( = \frac{\xi^{\alpha}}{\xi^{-1} - i \xi^0} \right) , \ \alpha = 1,2,3,4 \ (z^2 \, \overline z^2 = (z\overline z)^2 = 1) \right\}
$$
\be
\label{eq24}
\Leftrightarrow \left\{ z = g_c (x) : {\bm z} = \frac{{\bm x}}{\omega (x)} \, , \ z_4 = \frac{1-x^2}{2\omega (x)} \, , \ \omega (x) = \frac{1+x^2}2 - ix^0 \right\} .
\ee
Note that if we set $x_4 = -ix^0$ then $g_c$ in (\ref{eq24}) can be viewed as an involutive conformal transformation of the (real) euclidean space $E$ into itself, its inverse being given by the same formulas:
$$
x = g_c (z) : {\bm x} = \frac{{\bm z}}{\omega (z)} \, , \ x_4 = \frac{1-z^2}{2\omega (z)} \, , \ \omega (z) = \frac{1+z^2}2 + z_4
$$
\be
\label{eq25}
\left( \omega (z(x)) = \frac1{\omega (x)} \right) , \ dz^2 = \frac{dx^2}{\omega^2 (x)} \, , \ dx^2 = \frac{dz^2}{\omega^2 (z)} \, .
\ee

In a relativistic quantum theory of a local field $\varphi (x)$ (with energy-momentum spectrum in the forward light-cone) the vector-valued Wightman distribution $\varphi (x) \vert {\rm vac} \rangle$ is the boundary value of an analytic vector-function, $\varphi (x + iy) \vert {\rm vac} \rangle$, holomorphic in the {\it forward tube} domain
\be
\label{eq26}
{\mathcal T}_+ = \left\{ x + iy : x,y \in {\mathbb R}^4 \, , \ y^0 > \vert {\bm y} \vert \left( = \sqrt{y_1^2 + y^2_2 + y^2_3} \right) \right\} .
\ee
As a consequence the 2-point vacuum expectation value $w(x_1 - x_2) = \langle \varphi (x_1)$ $\varphi (x_2) \rangle_0$ admits an analytic continuation $w (x+iy)$ in the {\it backward tube} ${\mathcal T}_-$ (i.e. for $y^0 < - \vert {\bm y} \vert $). Remarkably, both tube domains ${\mathcal T}_{\pm}$ are conformally invariant -- without any assumption of conformal invariance of the underlying theory\footnote{Vladimir Glaser was also aware of this fact (private communication to the author of 1962).} \cite{U}. Noting further that $g_c$ acts without singularity on the tube domain (\ref{eq26}) we can surmise that one can define a compact picture transform $\phi (z)$ of the field $\varphi (x)$ such that the vector function $\phi (z) \vert {\rm vac} \rangle$ is analytic in the $g_c$-image of the tube domain (\ref{eq26}). In particular, it will admit a Taylor expansion in $z$, convergent in a neighbourhood of $z=0$.

\bigskip

\noindent {\bf Proposition 2.1.} {\it The vector valued function $\phi (z) \vert {\rm vac} \rangle$ is analytic in the image $T_+$ of ${\mathcal T}_+$ under the map $g_c$} (\ref{eq24}):
\be
\label{eq27}
T_+ = \left\{ z \in E_{\mathbb C} : \vert z^2 \vert < 1 \, , \ 2 \, z \, \overline z < 1 + \vert z^2 \vert \right\}.
\ee
{\it The closure $\overline M$ of the precompact image of the real Minkowski space $M$ in $E_{\mathbb C}$ has, in accord with} ({\ref{eq24}), {\it the form
\be
\label{eq28}
\overline M = \left\{ z = e^{2\pi i t} \, u \, ; \ t \in {\mathbb R} \, , \ u \in {\mathbb R}^4 \, , \ u^2 = {\bm u}^2 + u_4^2 = 1 \right\} .
\ee
The maximal compact subgroup $S(U(2) \times U(2))$ of $SU(2,2)$ acts on $z = (z_{\alpha})$ by euclidean rotations of ${\rm Spin} (4) \simeq SU(2) \times SU(2)$ and by multiplication with a $U(1)$ phase factor. Thus, it leaves the origin, $z=0$, invariant. Noting further the transitivity of the action of $SU(2,2)$ on $T_+$ we conclude that the (forward) tube is a quotient of the conformal group by its maximal compact subgroup:}
$$
T_+ = \frac{SU(2,2)}{S(U(2) \times U(2))} \, .
$$
(See Proposition 4.2 and Eq. (4.31) of \cite{NT}.)

\bigskip

The complex conjugation $x + iy \to x-iy$ in $M_{\mathbb C}$ interchanges the forward and backward tubes ${\mathcal T}_{\pm}$. It corresponds to the involution
\be
\label{eq29}
z \to z^* = \frac{\overline z}{\overline z^2} \, , \ T_{\pm}^* = T_{\mp} \, ,
\ee
which leaves compactified Minkowski space $\overline M$ pointwise invariant.

\smallskip

Since $g_c$ is a (complex) conformal transformation, in a conformal field theory the $x$- and the $z$-space correlation function will have the same expression. In particular, for a massless scalar field $\varphi (x)$ (of scale dimension 1) the corresponding $z$-picture field $\phi (z) = \frac{2\pi}{\omega (z)} \, \varphi (g_c (z))$ will have 2-point function
\be
\label{eq210}
\langle \phi (z_1) \, \phi (z_2) \rangle_0 = \frac1{z_{12}^2} \, , \ z_{12} = z_1 - z_2
\ee
holomorphic for $z_{12}^2 \ne 0$ (including the backward tube $z_{12} \in T_-$).

%%%
\section{Conformal vertex operators and massless fields}\label{sec3}
\setcounter{equation}{0}

The preceding discussion (Sect. 2) reflected (and updated) author's work of the mid 1980's \cite{T86}. Shortly afterwards Borcherds developed his elegant approach to 2-dimensional (chiral) vertex algebra \cite{B}. Nearly two decades later Nikolov extended it \cite{N} (on the basis of \cite{T86,NT01}) to higher dimensions. Here are some highlights of the resulting theory. (For a more detailed but still concise review see Sect. 4.3 of \cite{NT}.)

\smallskip

The state space of the theory is a (pre-Hilbert) inner product space ${\mathcal V}$ carrying a (reducible) vacuum representation of the conformal group $SU(2,2)$ such that:

\medskip

\noindent (a) The spectrum of the conformal Hamiltonian $H$ belongs to $\left\{ 0, \frac12 , 1 , \frac32 , \ldots \right\}$ and has a finite degeneracy:
\be
\label{eq31}
{\mathcal V} = \bigoplus_{\rho = 0,\frac12 , 1 , \ldots} {\mathcal V}_{\rho} \, , \ (H-\rho) \, {\mathcal V}_{\rho} = 0 \, , \ \dim {\mathcal V}_{\rho} < \infty \, .
\ee
Each ${\mathcal V}_{\rho}$ carries a fully reducible representation of ${\rm Spin} \, (4) = SU(2) \times SU(2)$. The central element $-\un$ of ${\rm Spin} \, (4)$ is represented by the {\it parity} $(-1)^{2\rho}$ on ${\mathcal V}_{\rho}$.

\medskip

\noindent (b) The lowest energy subspace ${\mathcal V}_0$ is 1-dimensional and is spanned by the (normalized) vacuum vector $\vert {\rm vac} \rangle$ which is invariant under the full conformal group $SU(2,2)$.

\medskip

The following proposition is a 4-dimensional counterpart of the state-field correspondence of 2D CFT (see Proposition 4.3 (c) of \cite{NT}):

\bigskip

\noindent {\bf Proposition 3.1.} {\it To every vector $v \in {\mathcal V}$ (of fixed parity $(-1)^{2\rho}$) there corresponds a unique local (Bose/Fermi for parity $+/-1$) field $Y(v,z)$ such that
$$
Y(v,0) \vert {\rm vac} \rangle = v \, , \ [H , Y(v,z)] = z \, \frac{\partial}{\partial z} \, Y(v,z) + Y(Hv,z)
$$
\be
\label{eq32}
[T_{\alpha} , Y(v,z)] = \frac{\partial}{\partial z_{\alpha}} \, Y(v,z) \, , \ \alpha = 1,2,3,4,
\ee
where $T_{\alpha}$ are the $z_a$-translation generators in the complexification ${\rm SL} (4,{\mathbb C})$ of $SU(2,2)$. If $v$ is a minimal energy state in an irreducible representation of $SU(2,2)$ then $v$ is annihilated by the special conformal generators $C_{\alpha}$ of the complexified conformal Lie algebra (since $[H,C_{\alpha}] = -C_{\alpha}$); such $v$ are called {\rm quasiprimary}.}

\bigskip

A prominent example of a quasiprimary state is the Fock vacuum $\vert 0 \rangle$ -- the lowest energy state of the massless scalar field $\phi (z)$ of 2-point function (\ref{eq210}) 
\be
\label{eq33}
\phi (0) \vert {\rm vac}\rangle = \vert 0 \rangle \ (= \vert 1,0,0 \rangle \ \mbox{-- i.e.,} \ (H-1) \vert 1,0,0 \rangle = 0 = {\rm spin} (4) \vert 1,0,0 \rangle) \, .
\ee
Note that the field vacuum $\vert {\rm vac} \rangle$ is $U(2,2)$ invariant -- i.e., it is annihilated by the full Lie algebra $u(2,2)$, while the Fock vacuum $\vert 0 \rangle$ (defined by $a \, \vert 0 \rangle = 0 = b \, \vert 0 \rangle$ (\ref{eq19})) is only annihilated by the lowering operators $F_i$ and by the generators $E_1 , E_3 , H_1 , H_3$ of $su(2) \oplus su(2)$. In the notation of Proposition 3.1 we can set $\phi (z) = Y(\vert 0 \rangle , z)$.

\smallskip

The statement that $\phi (z) \vert {\rm vac} \rangle$ is an analytic vector valued function for $z$ in the image $T_+$ (\ref{eq27}) of the forward tube is not obvious. In fact, the conjugation $z \to z^* = \frac{\overline z}{\overline z^2}$, which leaves invariant  the closure $\overline M$ of Minkowski space, transforms, according to (\ref{eq29}), $T_+$ into $T_-$. The hermitian on $\overline M$ field has a non-trivial law under conjugation for $z \in T_+$:
\be
\label{eq33bis}
\phi (z)^* = \frac1{\overline z^2} \, \phi (z^*) \quad \left({\rm as} \ (dz^*)^2 = \frac{d \, \overline z^2}{(\overline z^2)^2} \right)
\ee
so that the norm square of $\phi (z) \vert {\rm vac} \rangle$, is finite in $T_+$ (\ref{eq27}):
$$
(0 <) \Vert \phi (z) \vert {\rm vac} \rangle \Vert^2 = \langle {\rm vac} \vert \phi (z)^* \phi (z) \vert {\rm vac} \rangle = \frac1{1-2z\overline z + z^2 \, \overline z^2} < \infty
$$
\be
\label{eq34}
{\rm for} \quad 2 \, z \overline z < 1 + z^2 \, \overline z^2 \, .
\ee

\noindent {\bf Remark 3.1.} Note that for ${\bm x} = 0$ ($= {\bm z}$) $z_4 = \frac{1+ix^0}{1-ix^0}$; if the image of $x^0 = i$ is $z_4 = 0$, for the complex conjugate point $x^0 = -i$, $z_4$ is infinite, so that the map from the ket Fock vacuum $\vert 0 \rangle = \phi (0) \vert {\rm vac} \rangle$ to its bra counterpart $\langle 0 \vert$ is rather non trivial. The transformation law (\ref{eq33bis}) is related to a complex conformal inversion.

\bigskip

\noindent {\bf Proposition 3.2.} {\it The vertex operator $Y (\vert 0 \rangle , z)$ can be expressed in terms of the annihilation and creation operators $a^{(*)}$ and $b^{(*)}$} (\ref{eq110}) {\it as:
\be
\label{eq35}
(\phi (z) =) \ Y (\vert 0 \rangle , z) = U^* \exp (a^* q \, z \, b^*) \exp \left( \frac1{z^2} \, b \, q^* \check{z} \, a\right) U \, , \ \check{z} = \frac{z}{z^2} \, ,
\ee
where $q = ({\bm q} , q_4)$ are the quaternion units:
\be
\label{eq36}
q_j = - i \sigma_j \ (=-q_j^*) \, , \ j = 1,2,3, \ q_4 = \sigma_0 \ (= 1_2 = q_4^*) \, ,
\ee
\be
\label{eq37}
U \vert {\rm vac} \rangle = \vert 0 \rangle \ (= \phi (0) \vert {\rm vac} \rangle) \, , \ \langle {\rm vac} \vert \, U^* = \langle 0 \vert \, , \ UU^* = \un \, .
\ee
The complex translation generators $T_{\alpha}$ of} (\ref{eq32}), {\it given by
\be
\label{eq38}
T_{\alpha} = a^* q_{\alpha} \, b^* \, , \ \mbox{satisfy} \ \ T^2 := \sum_{\alpha = 1}^4 T_{\alpha}^2 = 0 \, .
\ee
The eigenspace ${\mathcal H}_n$ of the conformal Hamiltonian $H$} (\ref{eq117}) {\it corresponding to eigenvalue $n (=1,2,\ldots)$ is $n^2$-dimensional and consists of all homogeneous harmonic polynomials of degree $n-1$.}

\bigskip

The {\bf proof} is based on direct computations verifying that the 2-point function (\ref{eq210}) is reproduced by the vacuum expectation value of the product $Y (\vert 0 \rangle , z_1) \, Y (\vert 0 \rangle , z_2)$. The isotropy of the 4-vector $T_{\alpha}$ (\ref{eq38}) follows from the identity
\be
\label{eq39}
\sum_{\alpha = 1}^4 q_{\alpha}^{A_1 \dot B_1} \, q_{\alpha}^{A_2 \dot B_2} = 2 \, \epsilon^{A_1A_2} \, \epsilon^{\dot B_1 \dot B_2} \quad (\epsilon^{12} = - \epsilon^{21} = 1 \, , \ \epsilon^{AA} = 0) \, .
\ee
It implies, in turn, the harmonicity of the homogeneous polynomials $(Tz)^k$ ($Tz = \Sigma \, T_{\alpha} \, z_{\alpha}$). In view of the manifest $SO(4)$ invariance of both the 2-point function and the operator $Y (\vert 0 \rangle , z)$ (\ref{eq35}), it is enough to reproduce (\ref{eq210}) for $z_1 = ({\bm 0} , 1)$, i.e. to verify the expansion in spherical harmonics
\be
\label{eq310}
(1 - 2 z_4 + z^2)^{-1} = \sum_{k=0}^{\infty} h_k (z_4 , {\bm z}^2) \quad ({\bm z}^2 = z_1^2 + z_2^2 + z_3^2)
\ee
for

$$
h_k (z_4 , {\bm z}^2) = \frac1{(k!)^2} \, \langle 0 \vert (b \, a)^k (a^* q \, z \, b^*)^k \vert 0 \rangle \, :
$$
$$
h_0 = 1 \, , \ h_1 = 2 \, z_4 \, , \ h_2 = 3 \, z_4^2 - {\bm z}^2 \, , \ h_3 = 4 \, z_4^3 - 4 \, z_4 \, {\bm z}^2 \, ,
$$
$$
\ldots , h_k = (k+1) \, z_4^k - \begin{pmatrix} k+1 \\ 3 \end{pmatrix} z_4^{k-2} {\bm z}^2 + 2 \begin{pmatrix} k+1 \\ 5 \end{pmatrix} z_4^{k-4} ({\bm z}^2)^2 
$$
\be
\label{eq311}
- 2 \begin{pmatrix} k+1 \\ 7 \end{pmatrix} z_4^{k-6} ({\bm z}^2)^3 + \ldots
\ee
The space ${\mathcal H}_{k+1}$ of homogeneous harmonic polynomials of degree $k$ span the representation space $\left( \frac k2 , \frac k2 \right)$ of ${\rm Spin} \, (4) = SU(2) \times SU(2)$ of dimension $(k+1)^2$.

\bigskip

\noindent {\bf Remark 3.2.} The $n^2$-dimensional eigenspace ${\mathcal H}_n$ of $H$ is isomorphic to the stationary bound-state space of the non-relativistic hydrogen atom of energy $E_n = - \frac{m \, \alpha^2}{2 \, n^2}$ (where $m = \frac{m_e \, m_p}{m_e + m_p}$ is the reduced mass of the electron-proton system, $\alpha \sim e^2$ is the fine structure constant).

\bigskip

The fact that the same structure -- a UPEIR of the conformal group -- appears in two seemingly unrelated problems: a relativistic massless quantum field theory and the description of the bound states of the non-relativistic hydrogen atom -- i.e. the quantum mechanical Kepler problem -- should make us stop and think. An attempt to uncover the underlying general structure (an euclidean Jordan algebra?!) is being made by Guowu Meng -- see \cite{M13} and references therein.

%%%
\section{Partition function, modular form, Planck's \\ black-body radiation formula}\label{sec4}
\setcounter{equation}{0}

Knowing the spectrum of the conformal Hamiltonian $H$ (\ref{eq117}) on the Fock space ${\mathcal F}_0$ of a scalar massless field, we can write down its partition function
\be
\label{eq41}
Z(\tau) = {\rm tr}_{{\mathcal F}_0} \, q^H \, , \ q = e^{2\pi i\tau} \, , \ {\rm Im} \, \tau > 0 \, .
\ee
With our normalization both $\tau$ and $H$ are dimensionless. To display their physical meaning we should set
\be
\label{eq42}
2\pi \, {\rm Im} \, \tau = \frac{h \, \nu}{k \, T} \, ,
\ee
thus including the energy unit $h \, \nu$ and the absolute temperature $T$ (multiplied by the Boltzmann constant $k$) in the definition of $\tau$. Recalling that each energy level for a Bose particle gives rise to a geometric progression and inserting the energy eigenvalues and their multiplicities we find
\be
\label{eq43}
Z(\tau) = \prod_{n=1}^{\infty} (1-q^n)^{-n^2} \, .
\ee
The energy meanvalue in a thermal state characterized by the parameter $q$ (\ref{eq41}) (related to the temperature by (\ref{eq42})) is given by the logarithmic derivative of the partition function:
\be
\label{eq44}
\langle H \rangle_q = \frac1Z \, {\rm tr} (Hq^H) = q \, \frac d{dq} \, \ln Z(\tau) \, .
\ee
Remarkably, for a special choice of the zero-point energy,
\be
\label{eq45}
E_0 = - \frac18 \, B_4 = \frac12 \, \zeta (-3) = \frac1{240}
\ee
(where $B_k$ are the Bernoulli numbers and $\zeta (1-2k) = - \frac1{2k} \, B_k$ are special rational values of the analytically continued Riemann $\zeta$-function), $\langle H \rangle_q$ coincides with the unique modular form of weight 4:
\be
\label{eq46}
\langle H + E_0 \rangle_q = G_4 (\tau) = -\frac18 \, B_4 + \sum_{n=1}^{\infty} \frac{n^3 \, q^n}{1-q^n} \, .
\ee
More generally,
\be
\label{eq47}
G_{2k} (\tau) = - \frac{B_{2k}}{4k} + \sum_{n=1}^{\infty} \frac{n^{2k-1} \, q^n}{1-q^n}
\ee
satisfy, for $k = 2,3,\ldots$, the {\it modular covariance} condition
\be
\label{eq48}
\frac1{(c\tau + d)^{2k}} \, G_{2k} \left( \frac{a\tau + b}{c\tau +d}\right) = G_{2k} (\tau) \quad {\rm for} \quad \begin{pmatrix} a&b \\ c &d \end{pmatrix} \in SL(2,{\mathbb Z}) \, .
\ee
For $k = 2,3,4,5$ the function $G_{2k}$ satisfying the covariance condition (\ref{eq48}) (or, equivalently, such that the $k$-differential $G_{2k} (\tau)(d\tau)^k$ is modular invariant) is unique up to an overall normalization -- for a review of modular forms by a great master of the field -- see \cite{Z}; for a summary -- see \cite{NT}, Sect. 3.

\smallskip

Apart from its mathematical appeal Eq. (\ref{eq46}) has a direct -- and important -- physical meaning: it reproduces the Planck law for the black-body radiation. We shall summarize in what follows the discussion of \cite{NT05} and of Sect. 7 of \cite{NT}.

\smallskip

In order to make contact with reality we shall substitute the unit sphere in the definition (\ref{eq24}) (or (\ref{eq28})) of $\overline M$ by a sphere of radius $R$ (``the radius of the universe'', $R \gg 1$). If we also perform a uniform dilation $x \to \frac x{2R}$, $z \to Rz (x,R)$, $P_0 \to RP_0 + \frac1{2R} \, K_0$ we find
$$
{\bm z} (x,R) = \frac{{\bm x}}{2\omega \left( \frac x{2R} \right)} , \ z_4 (x,R) - R = \frac{ix^0 - \frac{x^2}{2R}}{2\omega \left( \frac x{2R} \right)} , \ 2 \, \omega \left( \frac x{2R} \right) = 1 + \frac{x^2}{4R^2} - i \, \frac{x^0}R
$$
\be
\label{eq49}
H_R = \frac{H(2R)}R = P_0 + \frac1{4R^2} \, K_0 \, .
\ee
It was observed over 40 years ago by Irving Segal that the universal cover of $\overline M$, the {\it Einstein universe} $\widetilde M = {\mathbb R} \times {\mathbb S}^3$, which admits a {\it global causal structure} is locally undistinguishible from $M$ for large $R$ (see for a concise expos\'e \cite{S82}). In particular, the Minkowski energy operator $P_0$ is well approximated, according to (\ref{eq49}) by the conformal energy $H_R$. In order to compare with the familiar expression for the black body radiation we restore the dimensional constants $h$ and $c$ setting $H_R = \frac{h c}R \, H(2R)$ -- instead of (\ref{eq48}) --  with the result:
\be
\label{eq410}
\langle H_R \rangle_q = \frac{h c}R \left( G_4 \left( \frac{ihc\beta}R \right) - E_0 \right) = \frac{hc}R \sum_{n=1}^{\infty} \frac{n^3 e^{-n \frac{hc\beta}{R}}}{1-e^{-n \frac{hc\beta}{R}}} \, ,  \ \beta = \frac1{kT} \, .
\ee
Each term in the infinite sum in the right hand side is a constant multiple of Planck's black body radiation formula for frequency
\be
\label{eq411}
\nu = n \, \frac cR \, .
\ee
Thus for a finite $R$ there is a minimal positive frequency $\nu_R = \frac cR$.

\smallskip

The expansion (\ref{eq410}) allows to derive in the limit $R \to \infty$ the Stefan-Boltzmann's law:
\be
\label{eq412}
{\mathcal E}_R (\beta) = \frac{G_4 \left( \frac{i\beta}{2\pi R} \right) - \frac1{240}}{R \, \nu_R} \underset{R \to \infty}{-\!\!\!-\!\!\!-\!\!-\!\!\!\longrightarrow} \frac{\pi^2 (kT)^4}{30 \, h^3 c^3} \, .
\ee

It also follows from the analysis of \cite{NT05,NT} that the thermal correlation functions are (doubly periodic) elliptic functions of periods $R/c$ and $i\beta$ (due to the Kubo-Martin-Schwinger boundary condition) \cite{NT05}.

\bigskip

The author thanks IHES for hospitality in December 2018 when this paper was written. He thanks Ludmil Hadjiivanov for his help in preparing the final version of the paper. His work was supported in part by contract DN-18/1 with the Bulgarian National Science Fund.

%\appendix
\section*{Appendix.\\ The ladder representation of $u(2,2) \subset sp (8,{\mathbb R})$}
\addcontentsline{toc}{section}{Appendix. The ladder representation of $u(2,2) \subset sp (8,{\mathbb R})$}
\label{apdxA}
\setcounter{equation}{0}

The aim of this Appendix is to give an elementary self contained exposition of the main result of \cite{MT} sketched in Sect. 1 above and in Sect. 3 of \cite{T10}.

\smallskip

It is convenient to use two realizations of the $\gamma$-matrices of ${\rm Cl} (4,1)$: the {\it Dirac picture} with diagonal $\beta = i \, \gamma^0$ and the {\it chiral picture} with diagonal chirality $\gamma_5$:
$$
\beta^D = \sigma_3 \otimes \un \equiv \begin{pmatrix} 
1 &0 &0 &0 \\
0 &1 &0 &0 \\
0 &0 &-1 &0 \\
0 &0 &0 &-1 \end{pmatrix} \ , \quad \gamma_5^D = \sigma_1 \otimes \un \quad\Rightarrow\quad 
$$
$$
\gamma_5^D \beta^D = \gamma_1^D \gamma_2^D \gamma_3^D = c^* \otimes \un\ ,\quad c = i \, \sigma_2 \ , \quad c^* = -c = \begin{pmatrix} 0 &-1 \\ 1 &0 \end{pmatrix} , \eqno ({\rm A.1D})
$$

$$
\beta^{\rm Ch} = \sigma_1 \otimes \un \ \ ( = \gamma_5^D) \ , \quad \gamma_5^{\rm Ch} = \sigma_3 \otimes \un \ , \quad \gamma_5^{\rm Ch} \beta^{\rm Ch} = \gamma_1^{\rm Ch} \gamma_2^{\rm Ch} \gamma_3^{\rm Ch} = c \otimes \un \ . \eqno ({\rm A.1Ch})
$$
They are related by a simple involutive similarity transformation:
$$
\gamma_{\mu}^D = V \gamma_{\mu}^{\rm Ch} V \  ,\ \ V^2 = \un \quad  ({\rm tr} \, V=0)\quad  \Rightarrow \quad \gamma_{\mu}^{\rm Ch} = V \gamma_{\mu}^D V \ ,
$$

$$
V = \frac1{\sqrt 2} \, (\sigma_1 + \sigma_3) \otimes \un \quad \Rightarrow \quad \gamma_1^{\rm Ch} \gamma_2^{\rm Ch} \gamma_3^{\rm Ch} = - \gamma_1^D \gamma_2^D \gamma_3^D \ . \eqno ({\rm A.2})
$$
A realization of $\gamma_j$, $j=1,2,3$ which obeys (A.1) and (A.2) is given by
$$
\gamma_j^{\rm Ch} = c \otimes q_j \equiv \begin{pmatrix} 0 &q_j \\ -q_j &0 \end{pmatrix} , \ \ q_j = -i \, \sigma_j \quad \Rightarrow \quad 
\gamma_j^D = -\gamma_j^{\rm Ch} \eqno ({\rm A.3})
$$
($q_j$ being the quaternion units $q_1 q_2 q_3 = -1 = q_j^2$).

\smallskip

The second quantized operators $\widehat X$ will be independent of the choice of basis if we transform simultaneously $\varphi , \widetilde\varphi$:
$$
\varphi^D = V \varphi^{\rm Ch} \quad\Leftrightarrow\quad \varphi^{\rm Ch} = V \varphi^D \quad\Leftrightarrow\quad \widetilde\varphi^D = \widetilde\varphi^{\rm Ch} V \, . \eqno ({\rm A.4})
$$
In accord with (\ref{eq16}) (\ref{eq17}) and (A.1) we shall set
$$
\varphi^D = \begin{pmatrix} a \\ b^* \end{pmatrix} \quad\Leftrightarrow\quad \widetilde\varphi^D = (a^* , -b) \ , \quad a_{\alpha} \vert 0 \rangle = 0 = b_{\alpha} \vert 0 \rangle \, , \ \alpha = 1,2 \, . \eqno ({\rm A.5})
$$
The CCR (\ref{eq14}) then imply
$$
[a_{\alpha} , a^*_{\beta}] = \delta_{\alpha\beta} = [b_{\alpha} , b_{\beta}^*] \ , \quad [a^{(*)}_{\alpha} , b_{\beta}^{(*)}] = 0 \ . \eqno ({\rm A.6})
$$
The {\it normal product}
$$
h :=\ :\!\widetilde\varphi \varphi\! : \ = \widetilde\varphi \varphi - \langle 0 \vert \, \widetilde\varphi \varphi \, \vert 0 \rangle \eqno ({\rm A.7})
$$
generates the centre $u(1)$ of $u(2,2)$ and will be identified with twice the helicity operator. Its spectrum is ${\mathbb Z}$ -- the set of all integers.

\smallskip

The entire  construction works for the Lie algebra $u(n,n) \subset sp (4n)$ of rank $2n$; Eqs. (A.5) (A.6) may be extended to $\alpha , \beta = 1,\ldots , n$. In general, one can introduce the Chevalley-Cartan basis of $sl (2n,{\mathbb C}) \supset su(n,n)$:
$$
E_i = \widetilde\varphi_i \, \varphi^{i+1} \, , \ F_i = \widetilde\varphi_{i+1} \varphi^i \, , \ H_i = [E_i , F_i] = \widetilde\varphi_i \, \varphi^i - \widetilde\varphi_{i+1} \, \varphi^{i+1} \, ,
$$
$$
i = 1,\ldots , 2n-1 \, . \eqno ({\rm A.8})
$$
For $\beta = \sigma_3 \otimes \un_n$ (which generalizes (A.1D)) one can define two Fock space representations -- one with vacuum defined by (\ref{eq17}) and another with the opposite convention
$$
E_+ \, \varphi \, \vert 0+ \rangle = 0 = E_- \, \widetilde\varphi \, \vert 0+ \rangle \, , \ E_- \, \varphi \, \vert 0- \rangle = 0 = E_+ \, \widetilde\varphi \, \vert 0- \rangle \, , 
$$
$$
E_{\pm} = {\textstyle \frac12} \, (1 \pm \beta) \, . \eqno ({\rm A.9})
$$
The representation $(h,+)$ with fixed eigenvalue $h$ of the invariant $: \widetilde\varphi \varphi :$ (A.7) and vacuum vector $\vert 0+ \rangle$ is a UPEIR with a {\it lowest weight} vector
$$
\vert h+ \rangle = \left\{ \begin{matrix}
N_1 \, \widetilde\varphi_n^h \, \vert 0 + \rangle \quad \ &{\rm for} &h \geq 0 \qquad\qquad\qquad\qquad \ \\ 
{ \ } \\
N_1 \, \varphi_{n+1}^{\vert h \vert} \, \vert 0+ \rangle &{\rm for} &h < 0 \, , \ N_1 = (\vert h \vert !)^{-1/2}
\end{matrix} \right. \eqno ({\rm A.10})
$$
satisfying, by definition,
$$
F_i \, \vert h + \rangle = 0 \, , \ i = 1,\ldots , 2n-1 \, , \ (H_{\theta} - 1) \, \vert h+ \rangle = 0 \eqno ({\rm A.11})
$$
where
$H_{\theta} = H_1 + \ldots + H_{2n-1}$ is the highest root. The representation $(h,-)$ with Fock vacuum $\vert 0-\rangle$ is a negative energy unitary irreducible representation with {\it highest weight} vector
$$
\vert h - \rangle = \left\{ \begin{matrix}
N_1 \, \varphi_n^{\vert h \vert} \, \vert 0 - \rangle \ &{\rm for} &h \leq 0 \\
{ \ } \\
N_1 \, \widetilde\varphi_{n+1}^h \, \vert 0 - \rangle &{\rm for} &h > 0
\end{matrix} \right. \eqno ({\rm A.12})
$$
such that
$$
E_i \, \vert h- \rangle = 0 \, , \ i=1,\ldots , 2n-1 \, , \ (H_{\theta} + 1) \, \vert h- \rangle = 0 \, . \eqno ({\rm A.13})
$$

The generator of the centre $u(1)$ of the maximal compact subalgebra $s(u(n) \oplus u(n))$ of $su(n,n)$ is given by
$$
H_c = H_1 + 2 H_2 + \ldots + n H_n + (n-1) H_{n+1} + \ldots + H_{2n-1} \, . \eqno ({\rm A.14})
$$
It is positive definite on $(h,+)$, negative definite on $(h,-)$ and satisfies
$$
(H_c - \varepsilon (\vert h \vert + n)) \, \vert h , \varepsilon \rangle = 0 \quad \mbox{for} \quad \varepsilon = \pm \, . \eqno ({\rm A.15})
$$
For the case of interest, $n=2$, and $\varepsilon = +$ we define the conformal Hamiltonian by (\ref{eq117}):
$$
H = {\textstyle \frac12} \, H_c = {\textstyle \frac12} \, (a^* a + b \, b^*) \quad \mbox{so that} \quad H \vert 0 \rangle = \vert 0 \rangle \ (= \vert 0+ \rangle) \, . \eqno ({\rm A.16})
$$

In order to recover the (Poincar\'e) energy-momentum we shall pass to the chiral representation (A.1Ch) of the Dirac algebra and set
$$
\varphi^{\rm Ch} = \begin{pmatrix} \overline\lambda \\ \partial \end{pmatrix} , \ \widetilde\varphi^{\rm Ch} = (- \overline\partial , \lambda) \, , \ \lambda = (\lambda_1 , \lambda_2) \, , \ \partial_A = \frac{\partial}{\partial \, \lambda_A} \, , \ A = 1,2 \, . \eqno ({\rm A.17})
$$
A straightforward calculation then gives:
$$
\widetilde\varphi \, \gamma_{\mu} \, \Pi_+ \, \varphi = i \, p_{\mu} \, , \ p_{\mu} = \lambda \, \sigma_{\mu} \, \overline\lambda \qquad \left( \Pi_{\pm} = {\textstyle \frac12} (1 \pm \gamma_5) \right). \eqno ({\rm A.18})
$$

The mass-shell condition $p^2 = 0$ (and Eq. (\ref{eq110})) then follow from the identity
$$
\sigma_0^{A \dot B} \, \sigma_0^{A' \dot B'} - {\bm\sigma}^{A\dot B} \, {\bm \sigma}^{A' \dot B'} = 2 \, \varepsilon^{AA'} \varepsilon^{\dot B \dot B'} \, . \eqno ({\rm A.19})
$$
Eqs. (A.2) (A.4) imply the relations
$$
a = \frac1{\sqrt 2} \, (\overline\lambda + \partial) \, , \ b = \frac1{\sqrt 2} (\lambda + \overline\partial) \, ,
$$
$$
a^* = \frac1{\sqrt 2} \, (\lambda - \overline\partial) \, , \ b^* = \frac1{\sqrt 2} \, (\overline\lambda - \partial) \, . \eqno ({\rm A.20})
$$
The lowest energy state (the zero helicity Fock vacuum) is given by
$$
\vert 0 \rangle = \frac2{\pi} \, e^{-\lambda \overline\lambda} \Rightarrow \langle 0 \mid 0 \rangle := \left( \frac2\pi \right)^2 \iint e^{-2\lambda\overline\lambda} \, d^2 \lambda_1 \, d^2 \lambda_2 = 1 \, . \eqno ({\rm A.21})
$$
The central element $H_c = H_1 + 2 H_2 + H_3$ (cf. (A.14)) assumes the form
$$
H_c = \lambda \overline\lambda - \overline\partial \partial \Rightarrow (H_c - 2) \, \vert 0 \rangle = 0 \, , \eqno ({\rm A.22})
$$
so that the spectrum of the conformal Hamiltonian $H = \frac12 \, H_c$ in the positive energy Fock space of the Heisenberg algebra of creation and annihilation operators coincides with the set $\left\{ 1 , \frac32 , 2 , \frac52 , \ldots \right\}$. The degeneracy of each eigenvalue of $H$ is determined by the corresponding representation of the spin group ${\rm Spin} (4) = SU(2) \times SU(2)$ generated by the vectors ${\bm J}_1 = \frac12 \, a^* {\bm \sigma} a \, , \ {\bm J}_2 = \frac12 \, b \, {\bm \sigma} \, b^*$ of spectrum ${\bm J}_i^2 = j_i (j_i + 1)$, $i=1,2$, $j_i = 0,\frac12 , \ldots$. In particular, the spectrum of $H$ in the zero helicity Fock space coincides with the set of positive integers and the degeneracy of the eigenvalue $n (= 2j+1)$ is $n^2$.

\end{document}